\documentclass[aps,reprint,floatfix]{revtex4-1}
\usepackage{amsmath}    
\usepackage{graphicx}  
\usepackage{verbatim}  
\usepackage{color}      
\usepackage{subfigure} 
\usepackage{hyperref}   
\usepackage{xcolor,txfonts}
\definecolor{SC-color}{named}{red}

\definecolor{SCh-color}{named}{green}

\begin{document}

\title{DC Josephson transport in a three-terminal Yu-Shiba-Rusinov system}
\author{Subrata Chakraborty}
\email[Correspondence to: ]{subrata@dubai.bits-pilani.ac.in}
\affiliation{Department of General Sciences, Birla Institute of Technology and Science, Pilani-Dubai Campus, Dubai International Academic City, Dubai 345055, UAE\looseness=-1}
  
\date{\today}

\begin{abstract} 
Scanning tunneling microscopy (STM) of magnetic adatoms on superconducting surfaces has established Yu-Shiba-Rusinov (YSR) states as a versatile platform for studying magnetic-superconducting interactions and phase-coherent quantum transport. Here, we investigate nonreciprocal supercurrent transport in a three-terminal Josephson junction comprising three BCS superconductors (`A', `B' and `C'), each coupled to a magnetic impurity hosting a pair of YSR states. Treating terminals `A' and `C' as primary transport electrodes and terminal `B' as a phase-control node, we demonstrate field-tunable nonreciprocal supercurrent between the primary electrodes. This effect requires both: the control terminal and broken particle-hole symmetry at least in one impurity attached to a primary electrode. Without the control terminal we show that the supercurrent remains reciprocal. Our results establish multi-terminal YSR junctions as a promising platform for engineering nonreciprocal superconducting transport and symmetry-breaking phenomena at the atomic scale.
\end{abstract}

\maketitle

\section{Introduction}
The development of scanning tunneling microscopy (STM) has opened new frontiers in probing the coupling between magnetism and superconductivity at the atomic scale, especially through experiments involving individual magnetic adatoms placed on superconducting surfaces. A hallmark signature of this coupling is the emergence of a pair of in-gap Yu-Shiba-Rusinov (YSR) bound states that form due to the magnetic impurity–superconductor interaction \cite{Yu1965, Shiba1968, Rusinov1969}.
In recent years, quantum tunneling and Josephson phenomena mediated by YSR states have attracted considerable attention, providing a versatile platform for exploring unconventional superconducting transport and many-body quantum effects \cite{Huang2020, Huang2021, Villas2021, Chakraborty2023}. Although this phenomena has been widely documented in mesoscopic devices—including superconductor-ferromagnet hybrids and quantum dots coupled to superconducting leads \cite{Ryazanov2001, Kontos2002, Rouco2019, Caruso2019}---its observation in STM-based platforms with single magnetic impurities has remained interesting. The Josephson effect, which underpins the coherent tunneling of Cooper pairs between superconductors, serves as a sensitive indicator of phase coherence. In standard Josephson junctions, the current-phase relation follows $I_S = \sum_n I_n \sin(n\chi)$, where $\chi$ is Josephson phase \cite{Rouco2019, Chakraborty2023}. This current-phase relation in the conventional Josephson junction is antisymmetric, $I_S(\chi)=-I_S(-\chi)$, due to time-reversal and inversion symmetries, and thus yielding reciprocal critical currents. Breaking these symmetries gives rise to nonreciprocal supercurrents, resulting in the Josephson diode effect (DE) \cite{Hu2007, Chen2018, Zhang2022, Misaki2021, Nikolic2026}. 
While external magnetic fields are frequently used to generate this nonreciprocity, significant attention is also directed toward field-free implementations \cite{DiezMerida2023, Bauriedl2022, Jeon2026, Gupta2023, Kokkeler2022, Zhang2024a, Nagata2025}. Magnetic adatoms deposited on one superconducting electrode provide an attractive route, as they naturally produce YSR states through spin-dependent scattering and can break key symmetries locally \cite{Trahms2023}. Furthermore, an imbalance in the normal-state density of states around the Fermi energy—i.e., broken particle-hole symmetry in the normal state (PHN)—can also provide an additional mechanism for generating nonreciprocal current \cite{Steiner2023, Ghosh2024}. By merging high-resolution YSR spectroscopy with phase-sensitive Josephson transport, this approach enables detailed exploration of parity-changing transitions and novel symmetry-breaking phenomena in magnetic impurity–superconductor systems down to the atomic limit.

In a different context, multi-terminal Josephson junctions—formed by connecting several BCS-type superconducting terminals—have attracted considerable attention in recent years \cite{Gupta2023, Chiles2023}. Their appeal largely stems from the rich, non-trivial topological properties that emerge in the synthetic phase space spanned by the superconducting phase differences \cite{Riwar2016, Strambini2016, Meyer2017}. Complementing this, recent experiments have successfully demonstrated nonreciprocal supercurrents in two-terminal Josephson devices by simultaneously tailoring time-reversal and particle-hole symmetries \cite{Ando2020, Narita2022, Jeon2022a}. These advances have inspired a broader range of studies examining nonreciprocal supercurrent behavior across diverse multi-terminal Josephson junction configurations, both with and without applied external magnetic fields \cite{Graziano2022, Yalcin2023, Cohen2018}.
\begin{figure}[h]
\begin{center}
\includegraphics[width=1\linewidth]{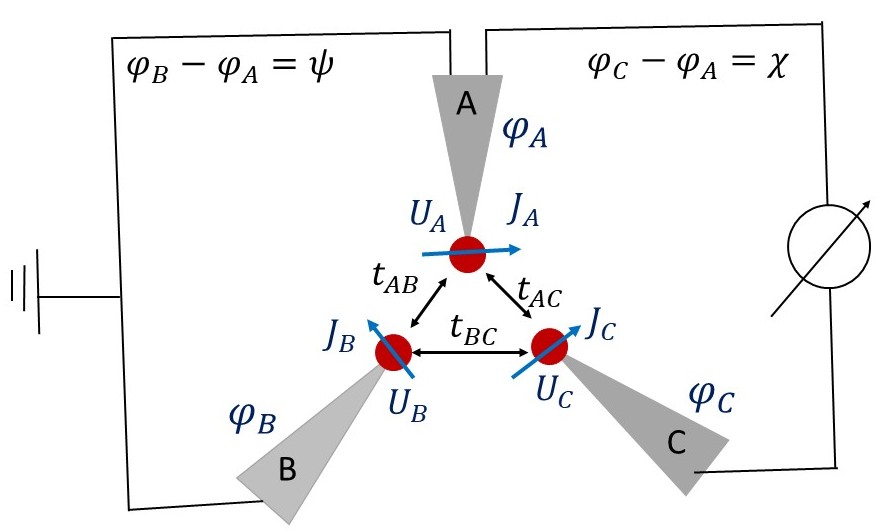}
\caption{\label{schematic}{Schematic of a three-terminal YSR system: Three magnetic impurities (red circles) are attached to three distinct superconducting terminals `A', `B' and `C'. The nonmagnetic energy and the exchange energy of a respective magnetic impurity are $U_{A/B/C}$ and $J_{A/B/C}$, respectively. The superconductor A/B/C is characterized by a superconducting phase $\varphi_{A/B/C}$. The spin-independent tunneling energy across A-B, A-C and B-C junctions are $t_{AB}$, $t_{AC}$ and $t_{BC}$, respectively. The A and B terminals are connected by a closed circuit; and the phase difference $\varphi_B-\varphi_A=\psi$ can be controlled by an external magnetic flux. The Josephson phase difference across the A-C junction is $\varphi_C-\varphi_A=\chi$.}}
\end{center}
\end{figure} 

In the present study, we investigate a three-terminal Josephson junction with three magnetic impurities, each magnetic impurity coupled to a distinct superconducting terminal, thereby hosting a pair of individual Yu-Shiba-Rusinov (YSR) states. Here, the three terminals `A', `B' and `C' are arranged in a cyclic quantum network, as schematically demonstrated in Fig.~\ref{schematic}. 
Next, considering `A' and `C' terminals as two electrodes we aim to investigate supercurrent flow between them. The phase difference between these two terminals can act as free variable. The remaining terminal `B' we can consider as the controlled-terminal, such that the phase difference between `B' and `A' terminals we can control via external magnetic field. We find that when particle-hole symmetry is broken at at least one of the magnetic impurity centers coupled to the superconducting electrodes, the presence of the control terminal can induce a nonreciprocal supercurrent between the two electrodes. Importantly, this asymmetry appears only in the presence of controlled-terminal; in its absence, a standard phase retains symmetric supercurrent.

The rest of this manuscript is organized as follows: In Sec. II, we describe the key model and Green function calculation to study the proposed nonreciprocal supercurrent physics. In Sec. III we present the numerical results and associated discussions. Finally, we conclude present work in Sec. IV.

\section{Model}
To study the supercurrent physics in a three-terminal YSR system we consider three magnetic impurities attached to three distinct superconducting STM tips `A', `B' and `C', as schematically represented in Fig.~\ref{schematic}. Each magnetic impurity hosts a pair of YSR states due to strong coupling with the respective superconducting tip. We further consider all the impurities are tunnel-coupled to each other causing YSR-hybridization. Superconducting tips are characterized by superconducting phases $\varphi_{\rm{A}}$, $\varphi_{\rm{B}}$ and $\varphi_{\rm{C}}$ for the tips `A', `B' and `C', respectively. We also assume that the terminals A and B are connected to a loop such that the superconducting phase difference between these terminals $\varphi_{\rm{B}} -\varphi_{\rm{A}}=\psi$ can be controlled by an external magnetic flux. In our theoretical set up here we are interested to investigate the supercurrent flow between the electrode terminals `A' and `C', where the phase difference between these two terminals we denote as $\varphi_{\rm{C}} -\varphi_{\rm{A}}=\chi$.

We now describe the three-terminal YSR system using an effective mean-field model Hamiltonian as follows
\begin{eqnarray}
\hat{H} &=& \hat{H}_{\rm{A}} +\hat{H}_{\rm{B}} +\hat{H}_{\rm{C}} 
+ \left(\hat{\bar{V}}_{\rm{AB}} + \hat{\bar{V}}_{\rm{BC}} + \hat{\bar{V}}_{\rm{AC}} +h.c \right), \label{jun25a1}
\end{eqnarray}
where $\hat{H}_j$ describes the subsystem formed by a magnetic impurity attached to the superconducting STM tip $j\in \left\{\rm{A,~B,~C}\right\}$ and $\hat{V}_{ij}$ represents the interaction potentials between subsystems $i$ and $j$ with $i,j\in \left\{\rm{A,~B,~C}\right\}$ and $i\neq j$. Considering the global spin-quantization along the $z$-axis we can choose the basis spinor for a magnetic impurity attached to the respective superconductor $j\in {\rm{A,~B,~C}}$ as $\tilde{d}_j^\dagger = \left( d_{j\uparrow}^\dagger,\; d_{j\downarrow},\; d_{j\downarrow}^\dagger,\; -d_{j\uparrow} \right)$. With respect to the global spin-quantization axis we further choose the basis spinor of the superconducting STM tip $j$ as $\tilde{c}_{{\bf{k}}j}^\dagger = \left( c_{{\bf{k}}j\uparrow}^\dagger,\; c_{-{\bf{k}} j \downarrow},\; c_{{\bf{k}}j\downarrow}^\dagger,\; -c_{-{\bf{k}} j \uparrow} \right)$. As a magnetic impurity attached to a superconductor hosts a pair of YSR bound states, hence in this work the Hamiltonian of a bare YSR system we can express as $\hat{H}_j = \hat{H}_{\mathrm{elec};j} + \hat{H}_{\mathrm{imp};j} +\hat{H}_{\mathrm{int};j}$ \cite{Chakraborty2023}, where 
\begin{align}
\hat{\bar{H}}_{\mathrm{imp}, j} &= \frac{1}{2} \, \bar{d}_j^\dagger \, \bar{H}_{\mathrm{imp}, j} \, \bar{d}_j,  \label{jun25a2} \\[6pt]
\hat{\bar{H}}_{\mathrm{elec}, j} &= \frac{1}{2} \sum_{{\bf{k}}} \bar{c}_{{\bf{k}} j}^\dagger \, \bar{H}_{\mathrm{elec}, {\bf{k}} j} \, \bar{c}_{{\bf{k}} j}, \label{jun25a3}  \\[6pt]
\hat{\bar{H}}_{\mathrm{int}, j} &= \frac{1}{2} \sum_k \left( \bar{c}_{k j}^\dagger \, \bar{H}_{\mathrm{int}, j} \, \bar{d}_j + \mathrm{h.c.} \right). \label{jun25a4}
\end{align}
In Eq.~\eqref{jun25a2} we define the Hamiltonian for a bare magnetic impurity attached to a superconductor $j$ with the Hamiltonian matrix $\bar{H}_{\mathrm{imp}, j} = U_j (\sigma_0 \tau_3) + \mathbf{J}_j \cdot (\boldsymbol{\sigma} \tau_0)$. The Hamiltonian for the corresponding superconductor is defined in Eq.~\eqref{jun25a3} with the Hamiltonian matrix $\bar{H}_{\mathrm{elec}, {\bf{k}} j} = \sigma_0 \left( \xi_{{\bf{k}} j} \tau_3 + \Delta_j e^{i \varphi_j \tau_3} \tau_1 \right)$. The interaction Hamiltonian between the magnetic impurity and the respective superconductor is defined in Eq.~\eqref{jun25a4} with the Hamiltonian matrix $\bar{H}_{\mathrm{int}, j} = v_j (\sigma_0 \tau_3)$. To describe the YSR systems we consider the coupling strength between a magnetic impurity and the corresponding superconductor as $v_j$. The Hamiltonian matrices in Eqs.~\eqref{jun25a2}-\eqref{jun25a4} are expressed in spin $\otimes$ Nambu space $\sigma_i\tau_j$, where $\sigma_i$ and $\tau_j$ stands for the Pauli matrices in the respective space. The electronic Hamiltonain of the superconductors are represented by electronic energy $\xi_{{\bf{k}} j}$, Cooper pairing energy $\Delta_j$, and superconducting phase $\varphi_j$. The bare magnetic impurities are defined by the single-particle nonmagnetic energy $U_j$ and the exchange field ${\bf J}_j$. Without loss of generality we set the exchange field of the STM tip-`A' along the global spin-quantization axis $z$ such that ${\bf J}_A =J_A~ (0,~0,~1)$, the exchange field for the tips-`B' as ${\bf J}_B =J_B~(\sin\tilde{\theta}\cos\phi,~\sin\tilde{\theta}\sin\phi,~\cos\tilde\theta)$ and the exchange field for the tips-`C'  as ${\bf J}_C =J_C~(\sin{\theta},~0,~\cos\theta)$ with $0 \leq \tilde{\theta}\leq \pi$, $0 \leq \phi \leq 2\pi$ and $0 \leq \theta \leq 2\pi$. The interactions between different YSR systems as appear in Eq.~\eqref{jun25a1} can be described by the tunneling Hamiltonian as follows \cite{Chakraborty2023}
\begin{align}
\hat{\bar{V}}_{ij} &= \frac{1}{2} \, \bar{d}_i^\dagger \, \bar{V}_{ij} \, \bar{d}_j, \label{jun25a5}
\end{align}
where $\bar{V}_{ij} = t_{ij} \, (\sigma_0 \tau_3) = \bar{V}_{ji}$,
and $t_{ij}$ is the hopping matrix element that describes the strength of spin-independent tunneling between the two respective YSR subsystems. 

In what follows, instead of working in the global spin-quantization frame we work in the mixed quantization frame. In the mixed frame each subsystem $j\in {\rm{A,~B,~C}}$ can be quantized along the respective exchange field ${\bf J}_j$, 
Upon quantization with respect a particular ${\bf J}_j$ direction corresponding subsystem can be described by retarded/ advanced ($r/a$) Green function as follows \cite{Chakraborty2023}
\begin{eqnarray}
\hat{g}_j^{r/a} (E) &=& \hat{g}_{j;\uparrow\uparrow}^{r/a}(E) \oplus \hat{g}_{j;\downarrow\downarrow}^{r/a}(E) \label{jun25a6} 
\end{eqnarray}
\begin{widetext}
with
\begin{align}
g^{r/a}_{j  \sigma \sigma}(E) 
= \frac{1}{D_{j\sigma}(E)}
\begin{pmatrix}
E \Gamma_j + \left(E + U_j - J_j \sigma \right)\sqrt{\Delta_j^2 - E^2} 
& \Gamma_j \Delta_j e^{i \varphi_j} \\[6pt]
\Gamma_j \Delta_j e^{-i \varphi_j} 
& E \Gamma_j + \left(E - U_j - J_j \sigma \right)\sqrt{\Delta_j^2 - E^2}
\end{pmatrix},  \label{jun25a7}
\end{align}
\end{widetext}
where $\Gamma_j = \pi N_{0,j} v_j^2$, ($N_{0,j}$ is the normal density of states of electrode $j$) and $D_{j\sigma}(E) = 2 \Gamma_j E (E - J_{j \sigma}) + \left[ (E - J_{j \sigma})^2 - U_j^2 - \Gamma_j^2 \right] \sqrt{\Delta_j^2 - E^2}$. 
The energy for the retarded/ advanced ($r/a$) Green function is defined as $E = E \pm i\gamma$ (with $\gamma \to 0^+$). In Eq.~\eqref{jun25a7} $J_{j\uparrow} = +J_j$ and $J_{j\downarrow} = -J_j$. The YSR bound states are obtained by solving $D_{j\sigma}(E) = 0$ for $J_j, \Gamma_j \gg \Delta_j$. And they satisfy $E_{\mathrm{YSR}, j \uparrow} = - E_{\mathrm{YSR}, j \downarrow}$.
with
\begin{align}
E_{\mathrm{YSR}, j \uparrow} 
= \Delta_j \,
\frac{J_j^2 - \Gamma_j^2 - U_j^2}
{\sqrt{\Gamma_j^2 + (J_j - U_j)^2} \;
 \sqrt{\Gamma_j^2 + (J_j + U_j)^2}}. \label{jun25a8}
\end{align}
The Green function as defined in Eqs~\eqref{jun25a6} and \eqref{jun25a7} for a bare YSR system can be decomposed as follows
\begin{align}
g^{r/a}_{j} &= g^{r/a}_{j;00} (\sigma_0  \tau_0) 
+ g^{r/a}_{j;03} (\sigma_0  \tau_3) 
+ g^{r/a}_{j;30} (\sigma_3  \tau_0) 
+ g^{r/a}_{j;33} (\sigma_3  \tau_3) \nonumber \\[6pt]
&\quad + 
f^{r/a}_{j;0+} (\sigma_0  \tau_+) 
+ f^{r/a}_{j;3+} (\sigma_3  \tau_+)  
+f^{r/a}_{j;0-} (\sigma_0  \tau_-) 
+ f^{r/a}_{j;3-} (\sigma_3  \tau_-) \label{jun25a9} ,
\end{align}
where $\tau_\pm= (\tau_1 \pm \tau_2)/2$ and $f^{r/a}_{j;l\pm} = f^{r/a}_{j;l} \, e^{\pm i\varphi_j}$.
The anomalous part, $f^{r/a}_{j;0}$ and $f^{r/a}_{j;3}$ correspond the singlet and mixed-triplet Cooper pair contributions to the corresponding bare sub-system's Green function. To study supercurrent physics we also need the tunneling matrices in the mixed quantization frame. With respect to the mixed quantization axes we can obtain the tunneling matrices as follows
\begin{eqnarray}
\hat{V}_{ij} &=& \hat{R}_i \hat{\bar{V}}_{ij} \hat{R}^\dag_j, \label{jun27a1}
\end{eqnarray}
where the rotational matrices for the `A', `B' and `C' subsystems are 
\begin{eqnarray}
\hat{R}_A &=& 1, \\
\hat{R}_B &=& \exp\left[{-i\frac{\phi}{2}\sigma_3}\right] \exp\left[{-i\frac{\tilde{\theta}}{2}\sigma_2}\right]\tau_0, \\
{\rm{and}}~~\hat{R}_C &=& \exp\left[{-i\frac{\theta}{2}\sigma_2}\right]\tau_0~,
\end{eqnarray}
respectively, with the exchange fields' orientation angles $0 \leq \tilde{\theta}\leq \pi$, $0 \leq \phi \leq 2\pi$ and $0 \leq \theta \leq 2\pi$, see previous texts.

\subsection{Supercurrent:} In this work our aim is to study supercurrent physics between the `A' and the `C' tips in presence of the `B' tip. Using the non-equilibrium Keldysh Green function formalism we can express the the DC Josephson current between `A' and `C' terminals as \cite{Chakraborty2023}
\begin{align}
& I_S^{AC} (\chi) = \frac{e}{2\hbar} \int_{-\infty}^\infty dE~{\rm{Tr}} \left[(\sigma_0\tau_3) \left\{\hat{V}_{AC}\hat{G}^{+-}_{CA}(E) - \hat{G}^{+-}_{AC}(E)\hat{V}_{CA} \right\}\right], \label{jun25a10}
\end{align}
where $\chi=\varphi_C -\varphi_A$, the tunneling matrices $\hat{V}_{ij}$ are defined in Eq.~\eqref{jun27a1}  and $\hat{G}^{+-}_{CA/AC} (E)=n_F(E) \left(\hat{G}^{a}_{CA/AC}(E) -\hat{G}^{r}_{CA/AC} (E)\right)$ is the Keldysh Green function component of the hybrid Green functions in the mixed quantization frame.
Here $n_F(E)$ is the Fermi-Dirac distribution at temperature $T$ and for chemical potential set to zero. In the the mixed quantization frame the hybridization between the `A' and the `C' tips' is accounted by the retarded/ advanced Green function $\hat{G}^{r/a}_{CA/AC}$. And the hybrid Green functions are obtained up to the second order perturbation correction as follows
\begin{eqnarray}
\hat{G}^{r/a}_{AC} &=& \hat{g}^{r/a}_A \hat{V}_{AC}\hat{g}^{r/a}_C + \hat{g}^{r/a}_A \hat{V}_{AB}\hat{g}^{r/a}_B \hat{V}_{BC}\hat{g}^{r/a}_C, \label{jun27a2} \\
\hat{G}_{CA} &=& \hat{g}^{r/a}_C \hat{V}_{CA}\hat{g}^{r/a}_A + \hat{g}^{r/a}_C \hat{V}_{CB}\hat{g}^{r/a}_B \hat{V}_{BA}\hat{g}^{r/a}_A. \label{jun27a3}
\end{eqnarray}
Next, using Eqs.~\eqref{jun25a10}-\eqref{jun27a3} we obtain simplified expression of the supercurrent between `A' and `C' tips as 
\begin{eqnarray}
\hspace{-7mm} I_S^{AC}(\chi,\psi) &=& t_{AC}^2\alpha\sin\chi + t_{AC}t_{AB}t_{BC} \left[(\beta+\lambda\cos\psi)\sin\chi \right. \nonumber \\
&& \left. -\lambda\sin\psi\cos\chi +\eta\sin\psi\right], \label{jun27a4}
\end{eqnarray}
where $\chi=\varphi_C-\varphi_A$ and $\psi=\varphi_B-\varphi_A$ are the phase differences between the corresponding tips, see previous texts. We here note that the phase difference $\psi$ we can control by an external magnetic flux-- a flexible knob. Following the Green function decompositions in terms of the spin $\otimes$ Nambu matrices for the YSR subsystems as shown in Eq.~\eqref{jun25a9}, we obtain the coefficients in Eq.~\eqref{jun27a4} as follows
\begin{eqnarray}
 \alpha &=& \frac{8e}{\hbar} \int_{-\infty}^\infty dE~ n_F(E)~ {\rm{Im}}\left[f_{A;0}^af_{C;0}^a +\cos\theta f_{A;3}^af_{C;3}^a\right], \label{jun27a5} \\
 \beta &=& \frac{8e}{\hbar} \int_{-\infty}^\infty dE~n_F(E)~{\rm{Im}}\left[\left(f_{A;0}^af_{C;0}^a
+ \cos\theta~ f_{A;3}^af_{C;3}^a\right)g^a_{B;03}\right. \nonumber \\
&& \left. + \cos\tilde{\theta}~ f_{A;3}^af_{C;0}^ag^a_{B;03} 
+ \cos(\theta -\tilde{\theta}) f_{A;0}^af_{C;3}^ag^a_{B;33} \right], \label{jun27a6} \\
 \eta &=& \frac{8e}{\hbar} \int_{-\infty}^\infty dE~n_F(E)~{\rm{Im}}\left[\left(f_{A;0}^ag^a_{C;03}
+ \cos\theta~ f_{A;3}^ag^a_{C;33} \right)f_{B;0}^a\right. \nonumber \\
&& \left. + \cos\tilde{\theta}~ f_{A;3}^af_{B;3}^ag^a_{C;03} 
+ \cos(\theta -\tilde{\theta}) f_{A;0}^af_{B;3}^ag^a_{C;33} \right], \label{jun27a7} \\
 \lambda &=& \frac{8e}{\hbar} \int_{-\infty}^\infty dE~n_F(E)~{\rm{Im}}\left[\left(f_{C;0}^ag^a_{A;03}
+ \cos\theta~ f_{C;3}^ag^a_{A;33}\right) f_{B;0}^a \right. \nonumber \\
&& \left. + \cos\tilde{\theta}~ f_{B;3}^af_{C;0}^ag^a_{A;33} 
+ \cos(\theta -\tilde{\theta}) f_{B;3}^af_{C;3}^ag^a_{A;03} \right]. \label{jun27a8}
\end{eqnarray}
We can now observe from Eq.~\eqref{jun27a4} that if at least one of the coefficients, $\lambda$ or $\eta$, is nonzero, then the system can exhibit nonreciprocal supercurrent for $\psi \neq 0$. On the other hand, from Eq.~\eqref{jun25a9}, we find that the absence of the nonmagnetic scattering potential ($U_j=0$) in a YSR terminal implies $g^{r/a}_{j;03}=g^{r/a}_{j;33}=0$. Consequently, Eqs.~\eqref{jun27a7} and \eqref{jun27a8} reveal that $\eta$ and $\lambda$ vanish for $U_C=0$ and $U_A=0$, respectively. Using Eqs.~\eqref{jun27a4}-\eqref{jun27a8} we can now investigate supercurrent physics in a three-terminal YSR system. 

\section{Results and discussions}
In what follows, we demonstrate nonreciprocal supercurrent flow in a three-terminal YSR system. We present numerical results based on the theoretical framework discussed in the previous section. We assume that all three superconductors are identical, with the temperature-dependent superconducting gap given by $\Delta_{\rm{A/B/C}}(T)=\Delta_0\,\tanh\!\left(1.74\sqrt{{T_c}/{T}-1}\right)$. The BCS superconducting gap at zero temperature is given by $\Delta_0 = 1.764\,k_B T_c$, where $T_c$ denotes the superconducting critical temperature. 
\begin{figure}[h]
\begin{center}
\includegraphics[width=0.8\linewidth]{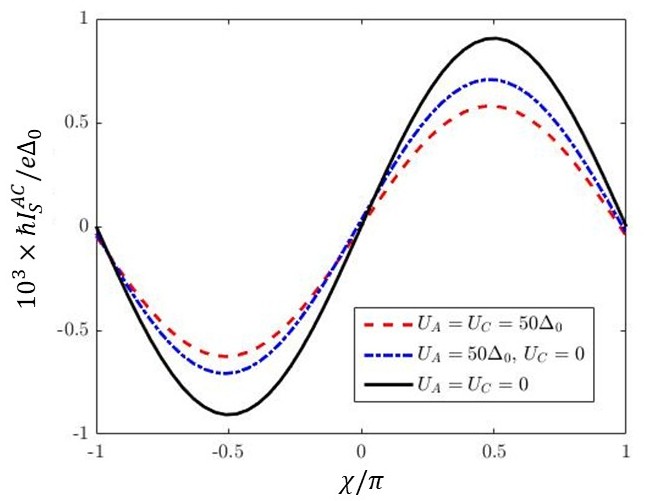}
\caption{\label{fig2}{Curren-phase relation $I_S^{AC}(\chi)$ across the A-C junction, setting $U_B=0$, $J_A=J_B=J_C=70\Delta_0$, $\Gamma_A=\Gamma_B=\Gamma_C=100\Delta_0$, $t_{AC}=t_{BC}=\Delta_0$, $t_{AB}=20\Delta_0$, $\theta=\pi/2$, $\tilde\theta=\pi/4$, $\psi=\pi/4$ and $T=0$.}}
\end{center}
\end{figure}

In Fig.~\ref{fig2} we illustrate the current-phase behavior between terminals `A' and `C' in the presence of an additional `B' terminal. This figure displays the dependence of the supercurrent $I^{AC}_s(\chi)$ on the superconducting phase difference $\chi$ across the A–C junction for different nonmagnetic energy $U_{A/C}$ of the A and C terminals. Figure~\ref{fig2} is obtained setting $U_B=0$, $J_A=J_B=J_C=70\Delta_0$, $\Gamma_A=\Gamma_B=\Gamma_C=100\Delta_0$, $t_{AC}=t_{BC}=\Delta_0$, $t_{AB}=20\Delta_0$, $\theta=\pi/2$, $\tilde\theta=\pi/4$, $\psi=\pi/4$ and $T=0$. The black solid line demonstrates the usual antisymmetric supercurrent behavior when $U_A = U_C = 0$, i.e., in presence of particle-hole symmetry both the electrode terminals `A' and `C'. In contrast, the blue dotted-dashed and the red dashed lines demonstrate the nonreciprocal current-phase behavior for the cases of $U_A = 50\Delta_0, U_C =0$; and $U_A =U_C =50\Delta_0$, respectively. We note that a nonzero nonmagnetic energy, $U_j \neq 0$, implies particle-hole symmetry breaking in the corresponding YSR system and, consequently, that $g^a_{j;03} \neq 0$ and $g^a_{j;33} \neq 0$. Hence, in Fig.~\ref{fig2} we can observe that a nonreciprocal (asymmetric) current (i.e., $  I^{AC}_S(\chi) \neq -I^{AC}_S(-\chi)  $) emerges whenever particle-hole symmetry is broken in at least one of the electrodes involved in the A–C transport path. In the symmetric case $U_A=U_C=0$--perfect particle-hole symmetry preserved in both `A' and `C' electrode terminals—the current-phase relation remains strictly antisymmetric, consistent with conventional reciprocity. In summary, here we find that once particle-hole symmetry is lifted in at least one of the electrodes—the remaining `B' terminal actively mediates the nonreciprocal supercurrent flow.

\begin{widetext}

\begin{figure}[h]
\begin{center}
\includegraphics[width=0.8\linewidth]{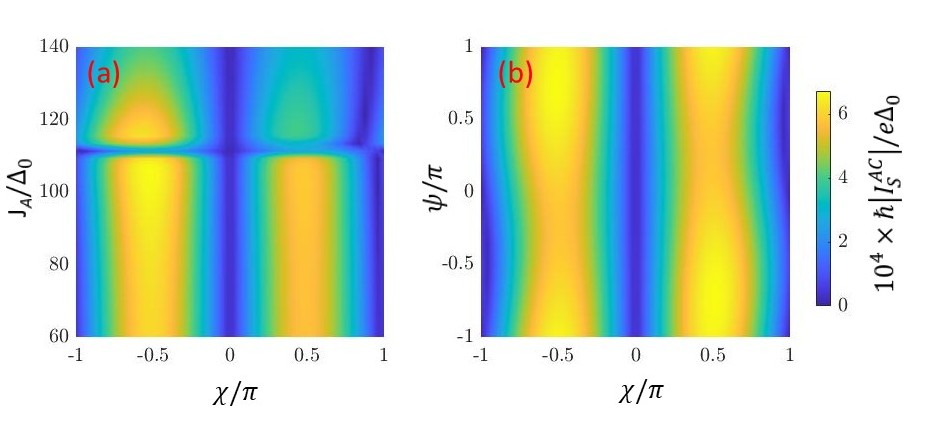}
\caption{\label{fig3}{Supercurrent amplitude across the A-C junction $\left|I_S^{AC}\right|$ (a) in $\chi$-$J_A$ space setting $\psi=\pi/4$; and (b) in $\chi$-$\psi$ space setting $J_A=70\Delta_0$. For both the panels $U_A=U_C=50\Delta_0$, $U_B=0$, $J_B=J_C=70\Delta_0$, $\Gamma_A=\Gamma_B=\Gamma_C=100\Delta_0$, $t_{AC}=t_{BC}=\Delta_0$, $t_{AB}=20\Delta_0$, $\theta=\pi/2$, $\tilde\theta=\pi/4$ and $T=0$.}}
\end{center}
\end{figure}

\end{widetext}

Next, in Fig.~\ref{fig3}(a) we plot the amplitude of supercurrent flow across the A-C junction $\left|I_S^{AC}\right|$ by varying the junction's phase difference $\chi$ and the exchange energy $J_A$ of terminal-`A'. Figure~\ref{fig3}(a) is obtained by setting $\psi=\pi/4$, $U_A=U_C=50\Delta_0$, $U_B=0$, $J_B=J_C=70\Delta_0$, $\Gamma_A=\Gamma_B=\Gamma_C=100\Delta_0$, $t_{AC}=t_{BC}=\Delta_0$, $t_{AB}=20\Delta_0$, $\theta=\pi/2$, $\tilde\theta=\pi/4$ and $T=0$. We notice a possible $0$-$\pi$ phase shift by varying the exchange energy $J_A$ in the supercurrent flow: This is similar to our earlier observation for the case of two-terminal YSR systems \cite{Chakraborty2023}. In Fig.~\ref{fig3}(a) for broken particle-hole symmetry at least in one of the electrode terminals, we further notice that the exchange energy $J_A$ also affects the extend of nonreciprocity in the current-phase relation $I_S^{AC}(\chi)$. We observe more pronounced nonreciprocity for larger $J_A$ values in Fig.~\ref{fig3}(a).

Finally, in Fig.~\ref{fig3}(b) we present the amplitude of the supercurrent $\left|I^{AC}_S\right|$ flowing through the A–C junction as a function of the phase difference $\chi$ and the control parameter $\psi$. Figure~\ref{fig3}(b) is plotted for a fixed $J_A=70\Delta_0$; and the remaining parameters are same as in Fig.~\ref{fig3}(a). The parameter $\psi$—phase difference between terminals `A' and `B'—can be controlled by varying external magnetic flux in the present theoretical set up. The two-dimensional color map in Fig.~\ref{fig3}(b) as well as the analytical Eq.~\eqref{jun27a4} clearly reveal a nonreciprocal current-phase relation in the A–C junction, satisfying $I^{AC}_S(\chi,\psi) \neq -I^{AC}_S(-\chi,\psi)$ for a fixed nonzero $\psi$. This asymmetry demonstrates the emergence of a nonreciprocal supercurrent controlled by the auxiliary phase $\psi$. In Fig.~\ref{fig3}(b) we further notice that, when the full synthetic phase space spanned by $(\chi,\psi)$ is considered, the supercurrent obeys the generalized antisymmetry relation $I^{AC}_S(\chi,\psi) = -I^{AC}_S(-\chi,-\psi)$, consider Eq.~\eqref{jun27a4} for more clarity. This property reflects an underlying symmetry of the multi-terminal system under simultaneous reversal of all superconducting phases, even though individual two-terminal subsets (such as A–C) exhibit broken reciprocity. The observed behavior underscores how the additional `B' terminal acts as a tunable knob that enables on-demand control of nonreciprocity in the primary A–C transport channel without compromising the global phase-reversal symmetry of the network. We note here that the parameter setting in Fig.~\ref{fig3}(b) corresponds to $\eta=\lambda$ in Eq.~\eqref{jun27a4}, which implies that $I_S^{AC}(\chi=0)=0$ for all values of $\psi$.

\section{Conclusions}
We have demonstrated the emergence of nonreciprocal supercurrent flow between two YSR qubits mediated by an additional YSR qubit in a three-terminal Josephson junction. We show that breaking particle-hole symmetry at a single magnetic impurity coupled to one of the electrode-terminals is sufficient to induce supercurrent nonreciprocity, thereby realizing a superconducting diode effect. Our results establish a simple and robust mechanism for engineering nonreciprocal Josephson transport in multi-terminal YSR systems and provide a promising platform for exploring Josephson physics and quantum functionalities in different YSR qubit architectures.

S.C. acknowledges the NFSG grant from BITS-Pilani, Dubai campus, which supported this research.


%

\end{document}